\documentclass[preprint]{aastex}

\def\citet#1{\citeauthor{#1} (\citeyear{#1})}

\def\citep#1{\citeauthor{#1}, \citeyear{#1}}

\def\referee#1{{#1}}

\begin{document}

\title{On a transition from 
solar-like coronae to rotation-dominated jovian-like magnetospheres
in ultracool main-sequence stars}

\author{Carolus J.\ Schrijver}
\affil{Lockheed Martin Advanced Technology Center,
3251 Hanover Street, Palo Alto, CA 94304}
\email{schrijver@lmsal.com}


\begin{abstract}

For main-sequence stars beyond spectral type M5 the characteristics of
magnetic activity common to \referee{warmer solar-like stars change} into the
brown-dwarf domain: the surface magnetic field becomes more dipolar
and the evolution of the field patterns slows, the photospheric plasma
is increasingly neutral and decoupled from the magnetic field,
chromospheric and coronal emissions weaken markedly, and the
efficiency of rotational braking rapidly decreases. Yet, radio
emission persists, and \referee{has been argued} to be dominated by
electron-cyclotron maser emission \referee{instead of the
gyrosynchrotron emission from warmer stars}.  These properties may
signal a transition in the stellar extended atmosphere. Stars warmer
than about M5 have a solar-like corona and wind-sustained heliosphere
in which the atmospheric activity is powered by convective motions
that move the magnetic field. Stars cooler than early-L, in contrast,
may have a jovian-like rotation-dominated magnetosphere powered by the
star's rotation \referee{in a scaled-up analog of the magnetospheres
of Jupiter and Saturn}.  A dimensional scaling relationship for
rotation-dominated magnetospheres by Fan et al.\ (1982) is consistent
with this hypothesis.

\end{abstract}


\section{Introduction}

Main sequence stars have a convective envelope around a radiative
core from about spectral type A2 (with 
an effective temperature of $T_{eff} \approx 9000$\,K) to about M3
($T_{eff} \approx 3200$\,K). Stars warmer than about F2 
($T_{eff} \approx 7000$\,K) have but a shallow
convective envelope that sustains at most weak magnetic activity. 
As one of these stars, the Sun exhibits many of the properties
of the magnetically-driven variability of the population of F2-M3
dwarf stars. The radiative losses from their chromospheres,
transition regions and coronae generally decrease with age as
stars lose angular momentum through a magnetized wind. The radiative
losses from these distinct thermal domains 
scale through power laws with the average magnetic flux
density on the stellar surface. All of these stars exhibit
signatures of flaring that increase with increasing quiescent activity. 
The radio emission from all of these stars, albeit generally weak,
scales lineary with the coronal X-ray emission, suggestive of persistent 
non-thermal energetic-particle populations within their outer atmospheres
(e.g., \citep{schrijver+zwaan99}, and references therein;
see \citep{berger+etal2008}, for a recent version of the radio/X-ray
scaling).

\referee{Stars cooler than about M2 to M4 are expected to be fully 
convective (\citep{chabrier+baraffe2000}), lacking the 
convective overshoot layer and
tachocline that are thought to be important to the solar dynamo. Yet
these stars are capable of generating magnetic field, with a
rotation-activity relationship that continues clearly to at least
spectral type M8 (e.g., \citep{mohanti+etal2002},
\citep{reiners+basri2009}).  
However, their} magnetic field appears to be a
predominantly axi-symmetrical large-scale poloidal field with at most
a slow evolution in the surface pattern (\citep{donati+etal2008}; also
\citep{reiners+basri2009}), in contrast to warmer stars that show
relatively-rapidly evolving, non-axisymmetric fields evolving subject
to flux emergence and differential rotation.

Spectra of  
FeH lines reveal that stars at least down to M9 have magnetic fields,
and that the coolest stars beyond M6 for which fields can be
measured have magnetic flux densities of $fB=1.5$\,kG or more (e.g.,
\citep{berger+etal2008a}, \citep{hallinan+etal2006} and 2007,
\nocite{hallinan+etal2007} \citep{reiners+basri2007}.  X-ray flaring
has been reported down to at least M9.5 (e.g.,
\citep{liebert+etal1999}, \citep{reid+etal99},
\citep{fleming+etal2003}, \citep{berger+etal2008}).

Beyond about M8, the relationship between
chromospheric (H$\alpha$) or coronal (X-ray) emission and rotation
rate becomes a weak tendency and even that disappears beyond
L0 (e.g., \citep{reiners+basri2009},
\citep{basri2008}; see \citep{mohanti+basri2003}, for
a discussion of difficulties in  assigning
spectral types). Whereas essentially
all stars of spectral type M8 have strong H$\alpha$ emission, this
drops to 60\%\ at L0, $\approx 15$\%\ at L4, and to less than
than $\approx 10$\%\ by L5 (e.g.,
\citep{mohanti+etal2002}, \citep{mohanti+basri2003},
\citep{west+etal2004}, \citep{reiners2007},
\citep{reiners+basri2007}, 
\citep{schmidt+etal2007}).

In contrast to the weakening of 
the rotation-activity relationship for the
traditional chromospheric (H$\alpha$) and coronal (X-ray)
indicators of magnetic activity beyond about M8, the radio
luminosity continues to increase with increasing angular velocity
up to at least L0 (\citep{berger+etal2008}), with (often variable) 
radio emission
detected to at least L3.5 (\citep{berger2006}). Interestingly, the
essentially linear relationship between radio and X-ray
luminosities for all other cool stars breaks down beyond M5, 
with the X-ray luminosity dropping by a factor of about
3,000 below that relationship at a given radio luminosity by spectral type L0 
(\citep{berger+etal2008}).

The reduction of the traditional diagnostics for chromospheric and coronal
activity, the weakening of their dependence on rotation rate
despite the presence of strong magnetic fields, and
the increase in time scale for rotational braking (discussed in Sect.~4),
may have their origin
in the very low degree of ionization of the photospheric plasma. 
\citet{mohanti+etal2002}
argue that
beyond M5, the photospheric
plasma is weakly coupled to the magnetic field, and the high
resistivity  and associated diffusivity
of the plasma should render it very difficult to generate
electrical currents by convection-driven field motions or to
transmit any generated in the
stellar interior to the corona
(see also \citep{mohanti+basri2003}, who also discuss the
role of 
dust in these cool atmospheres).
The persistence of radio emission into the L-type range, despite
these changes in activity characteristic of all other cool stars,
suggests we
explore a fundamental change in the character of stellar magnetic
activity: the weakening 
of a well-developed convection-powered stellar corona and
associated stellar wind may cause the stellar outer atmosphere to
change from a solar-like corona with heliosphere to a Jupiter-like
magnetosphere shaped by the incoming wind of the interstellar medium
(ISM) 
impinging on the stellar magnetic field, while powered by the stellar rotation.

\section{Radiative and rotational energy losses}
Coronal emission and magnetic braking  both weaken
rapidly beyond mid-M. \referee{Up to about spectral
type L0, the energy lost by rotational braking appears to be consistent with 
that expected from a solar-like coronal domain
beneath an outflowing wind, 
as can be seen from the
following argument.} I
start from the premise that the stellar atmosphere is 
essentially  hydrostatically stratified. Near the stellar surface,
an isothermal plasma has a pressure scale height
\begin{equation}
H_{\rm p}\,=\,{akTm_p R_\ast^2 \over G M_\ast},
\end{equation}
with stellar radius $R_\ast$, temperature $T$, proton mass $m_p$,
and gravitational constant $G$. Here, I assume a pure hydrogen gas
for an  order-of-magnitude estimate. The constant $a=1$ for a 
neutral gas, and $a=2$ for fully-ionized hydrogen. 
A largely-neutral photospheric gas of $T=2000$\,K on a compact ultra-cool
star with surface gravity $g\sim 10^{5.4}\sim 9 g_\odot$\,cm/s$^2$ 
would have $H_p(2000\,K)\sim 9$\,km.

In order to lead to magnetic braking, plasma of sufficient density
needs to exist out to the lesser of the Alfv{\'e}n radius
(to which corotation of star and high-atmospheric plasma is enforced) and the 
radius of synchronous rotation (beyond
which centrifugal forces dominate). The latter (which
is likely the smaller of the two for rapidly-rotating
L-type dwarf stars) is 
\begin{equation}
R_S = \left( {GM_\ast \over \Omega^2}\right )^{1\over 3},
\end{equation}
for stellar mass $M_\ast$ and angular velocity $\Omega=2\pi/P$, at
rotation period $P$. The characteristic 
value of $R_S\sim 7R_\ast$ (see Table~1) 
is so much larger than $H_p(2000\,K)$ that the plasma density
at $R_S$ would be
insufficient to lead to significant angular momentum loss. 

For a plasma at coronal temperatures the density scale height is
much higher, of course.
Only very few stars beyond  M9 have been detected
in X-rays (e.g., \citep{audard+etal2007},
\citep{robrade+schmitt2008}). Their quiescent emissions
are at a level of about $L_X/L_{\rm bol}\sim 10^{-4}$, or 
$\log(L_X)\sim 25.5$, while others have even lower upper limits. 

Assuming a hydrostatically stratified
atmosphere, the electron density $n_e$ associated
with such an emission can be estimated from
\referee{the X-ray luminosity $L_X$ based on the
plasma's volumetric emission, $n_e^2 \Lambda(T_C)$, and 
characteristic
volume, $4\pi R_\ast^2 H_p(T_C)$,}
\begin{equation}\label{eq:densityestimate}
L_X =  4\pi R_\ast^2 H_p(T_C) \, n_e^2 \Lambda(T_C)
\end{equation}
for a plasma at temperature $T_C$, with an emissivity
$\Lambda(T_C)\approx 2\times 10^{-18}T^{-2/3}$ (for $\log(T_C)\in [5.5,7.5]$); 
for $T_C=1.5$\,MK,
comparable to that of the bulk of the solar wind and
consistent with the very-limited  X-ray spectral information available on
L-type dwarf stars (\citep{robrade+schmitt2008}), 
$\Lambda(1.5\,{\rm MK})\sim 10^{-21.8}$ (e.g., \citep{schrijver+zwaan99}). 
The coronal-base density thus
estimated for L-type stars with $\log(L_X)\sim 25.5$ and 
$T_C=1.5$\,MK is $n_0 \sim 10^{8.8}$\,cm$^{-3}$.
 
To estimate the plasma density at height $R_S$ within the 
equatorial plane, the density 
for an isothermal atmosphere can be approximated 
from a balance between pressure gradient, gravity, 
and centrifugal acceleration:
\begin{equation}\label{eq:strat}
{{\rm d}n(r) \over {\rm d}r} = - \left(GM_\ast m_p \over 2kT_C \right)
{n(r) \over r^2} + \left({\Omega^2 m_p \over 2kT_C}\right) n(r)\,r
\end{equation}
(note that the stratification of a subsonic wind is very similar
to a static stratified atmosphere, e.g., \citep{hpI-9}). 
With the above value of $n_0$, $n(R_S) \approx
10^{5.6-5.8}$\,cm$^{-3}$ for the characteristic L2 and L5 stars
in Table~1  (the centrifugal force below  $R_S$ modifies the density
profile by a factor of $\la 2$ relative to the stratification
in the absence of rotation).

Now assume that the centrifugal force beyond $R_S$
accelerates plasma outward, which is replenished from
below at (at most) the thermal velocity
\begin{equation}
v_{\rm th} = \left (  \gamma {kT_C \over m_p} \right ) ^{1 \over 2},
\end{equation}
for adiabatic index $\gamma$, so that the upper limit for the 
mass loss is 
\begin{equation}
{\dot M} = \alpha \,4\pi R_\ast^2\,v_{\rm th}\,m_p\,n(R_S),
\end{equation}
with $\alpha$ a geometry factor. If the mass loss were 
isotropic, $\alpha\equiv 1$, but as mass is probably 
lost from only part of the surface area, I use  
$\alpha=1/2$ below.

The stellar magnetic field can enforce corotation out to
a distance $R_A$ where the plasma $\beta$ becomes of order unity, given by
\begin{equation}
{B_0^2 \over 4\pi} \left( {R_\ast \over R_{A}}\right) ^6 = 2n(R_{A})kT_C,
\end{equation}
if the field is approximated by a 
dipole of characteristic strength $B_0$ at the
stellar surface; $n(r)$ is assumed to be 
given by Eq.~(\ref{eq:strat}).

Outflowing plasma in a spherical shell carries an angular momentum
\begin{equation}
{\dot L} = {2 \over 3} \Omega R_A^2 {\dot M},
\end{equation}
and a rotational energy of
\begin{equation}\label{eq:wc}
W_C = {\dot L}\Omega.
\end{equation}
With the values in Table~1, one finds $W_C\approx 10^{25.8}$\,erg/s
for $\log(L_X)\sim 25.5$.
For the characteristic L2 star in Table~1, 
this compares well to the estimated loss $W_{\rm rot}$
from rotation-age studies (see Sect.~4), \referee{so that
we can conclude that the existence of a faint thermal corona in early L-type
stars is compatible with the inferred magnetic braking.}

The estimatd $W_C$ is while about an order of magnitude larger 
than $W_{\rm rot}$ for the M7 dwarf star. This  
may be because of the exponential dependence $W_C$ on
$T_C$, or 
because the corona of the M7 star may be solar-like, in the
sense that a few relatively bright and dense regions dominate the emission,
so that the base density from Eq.~(\ref{eq:densityestimate}) 
would be an overestimate for the highest coronal loops. 

The estimates in this section obviously  depend on the
assumed values of $L_X$ and $T_C$. The
value of $W_{\rm C}$ scales with $L_X$ about as  
a square root, so that our conclusions are not strongly modified
if the characteristic value of $L_X$ observed for the L-dwarf
binary Kelu-1 is somewhat larger than what is characteristic of
the ensemble of similar L-type dwarf stars. The dependence on $T_C$ 
is much stronger, not surprisingly, 
because of the near-exponential stratification of density in
units of the pressure scale height: for $T_C=1-3$\,MK, 
$\log(W_{\rm rot})=24.2-27.3$.

\section{The magnetosphere of the coolest dwarf stars}
Even though magnetic braking remains compatible with the concept
of a hot corona, the radio emission in ultracool dwarfs is
disproportionally strong. One possible cause for this that should
be explored, and can then be ruled out, is the interaction with the ISM.
Observations suggest that there is relatively little 
heated coronal plasma well into the L-type spectral range, 
there should not be much of an associated stellar
wind either. The motion of a star with at most a weak wind
relative to the ISM 
should lead to the formation of a magnetopause. 
\referee{The energy input into the stellar
magnetosphere from this interaction is expected to be low, 
based on the following.}

The standoff (or Chapman-Ferraro) distance of the magnetopause 
is set by where the \referee{dynamic pressure 
$\rho_w v_w^2$ of the ISM wind with density $\rho_w$ and 
relative velocity
$v_w$ equals} the magnetic
pressure of the stellar field. 
Observations suggest that the large-scale field of
very cool dwarf stars can be approximated by a dipole field, so that the field
scales with $\mu /r^3$, with the magnetic moment
$\mu=B_\ast R^3_\ast$, where $B_\ast$ is the
field strength near the stellar equator. $R_{CF}$
for a late M-type or L-type
dwarf star can thus be estimated 
\referee{from $\rho_w v_w^2 \propto B^2(R_{CF})/8\pi$, or:}
\begin{equation}\label{eq:cfstar}
{R_{CF} \over R_\ast} \sim 3.5 \bigl( {B_G^2 \over n_0 v^2_{100}}
\bigr)^{1/6}.
\end{equation}
The constant of proportionality is determined by using the observation
that for Earth $R_{CF,\oplus}\approx 10$\,Earth radii  
\referee{(\citep{russell2007})}, for a wind speed of $v_{100}\sim 4$ 
(in units of
100\,km/s) and a characteristic solar-wind particle density
$n_0=10$\,cm$^{-3}$ \referee{(\citep{feldman+etal1977})}, and $B_G \sim
0.6$\,G for the Earth's polar field strength \referee{(\citep{allen72})}.

The relative motion of stars through the ISM averages 
about 40\,km/s for a sample of very cool 
dwarf stars (e.g., \citep{wood+etal2005}, \citep{schmidt+etal2007}). 
For a hypothetical dwarf star that has  no stellar
wind, that moves through an ISM like that around the
solar system with $n_0=0.1$ (e.g., \citep{wood+etal2002}) 
at $v_{100}=0.4$, and $B_G=1.5$\,kG, one finds 
${R_{CF,\ast}/R_\ast} \approx 32$; that
relative scale is similar to the geometry in the case of Jupiter's
magnetosphere, for which $R_{MP}/R_{\rm J}=42$
(e.g., \citep{walker+russell1995}).

To estimate how much power the ISM wind might impart onto a
stellar bowshock and potentially into the stellar magnetosphere, we
estimate the power in the bulk kinetic energy 
over the cross section of the bow shock with cross section $\pi R^2_{CF}$:
\begin{equation}
P_w\approx 3\times 10^{20} n_0 v_{100}^3
\left( {R_{CF}\over R_\ast}\right)^2
\left( {R_\ast \over R_\odot}\right)^2,
\end{equation}
so that at $n_0=0.1$, $v_{100}=0.4$, and $R_\ast=0.3R_\odot$,  
$P_w\approx 10^{20-22}$\,erg/s for characteristic M7 to L5 stars. 
This lies orders
of magnitude below the H$\alpha$ and X-ray luminosities
(see Table~1) and below the power that needs to be extracted from the stellar
rotation with age (see below). Hence, 
the ISM wind is not a significant source of magnetospheric
activity in dwarf stars through mid-L, even
if it shapes a close-in asteropause.

\section{From a stellar corona to a planetary magnetosphere}
As the solar-like
corona fades away from mid-M to mid-L, 
one may expect
increasing signatures of a rotation-dominated magnetosphere towards
late L-type stars like
that for Jupiter. 
In such a magnetosphere, energy is taken from 
star's rotational energy through a torque applied by outflowing
plasma. 
The energy loss, $W_{\rm rot}$, can be estimated from
the rotation-age relationship.
\citet{reiners+basri2008} infer time scales
for magnetic braking that increase from about 3\,Gyr for an M7
star to 7\,Gyr for an L2 star, and over 10\,Gyr
for an L5 star (the latter is relatively
poorly constrained and, perhaps, much larger). Using
\begin{equation}
W_{\rm rot} = {{\rm d} \over {\rm d}t} {1 \over 2} I \Omega^2
\end{equation}
yeidls  $W_{\rm rot}\sim 
10^{26}$\,ergs/s (Table~1). 
In the early L-type range, this is comparable to $W_C$ 
from Eq.~(\ref{eq:wc}) that could be taken away by the outflow from
the relatively weak corona, i.e., that would be plausibly consistent with
the existence of such a weak corona. But both $W_{rot}$ and $W_C$ 
exceed X-ray and H$\alpha$ losses, i.e., 
the rotational energy loss exceeds the radiative
losses, in contrast to what is seen in warmer solar-type stars. 

Let us now explore an order-of-magnitude scaling for the 
power expected from a rotationally-dominated magnetosphere. 
\citet{fan+etal1982} argue for a scaling based on a dimensional
analysis ('principle of similitude') that relates the
magnetic moment, $M_B=B\ell^3$, the rotation period, $P$, and the 
radius, $R_a$, at which the acceleration process begins to the 
power, $W^\prime$, generated:
\begin{equation}\label{eq:fan}
W^\prime = {4\pi^2 K \over c^2} {M_B^2 \over P^2 R_a},
\end{equation}
where the constant $K$ has to follow from a measurement until a
full theory is developed. 
Based on the properties in
Table~\ref{tab:table}, using a mean photosperic field strength
of $1.5$\,kG, and with $R_S$ for $R_a$, values are found for
$^{10}\log(K\cdot W_{rot}/W^\prime)$ of $0.6$, $0.4$, and $0.6$  
for the characteristic M7, L2, and L5 stars,
respectively. \citet{fan+etal1982} estimate a range of
values for the constant $^{10}\log(K)$ for Jupiter from $-1.7$ to $-0.7$.
The value of $K$ for Jupiter lies, 
remarkably, within the range of values needed to let
$W_{rot}$ and $W^\prime$ be comparable
for the stars in Table~1. 
In contrast, this scaling  applied to the Sun
yields $\log(W^\prime_\odot) \sim 21.3-22.3$, which is 
well over a thousand times less. 
At least $W^\prime$ for ultracool dwarfs
is much larger than the solar value, 
while with the only available
calibration point -~Jupiter~- one concludes that $W^\prime$ is close to both
$W_{rot}$ and $W_C$ for an early L-type star.  

\section{Discussion and conclusions} 
Somewhere along the spectrum of
stars, brown dwarfs, and \referee{planets} a transition from an
outflow-driven asterosphere to a field-shielded rotating magnetosphere
must occur. Based on the observational evidence, I argue that this
transition occurs at the bottom end of the true stellar range of the
main sequence: for stars cooler than about spectral type M5, the
properties characteristic of solar-like activity progressively
disappear, while, in contrast to warmer stars, magnetic braking
extracts more energy from the star than needed to power the
chromospheric H$\alpha$ and (weak) coronal X-ray emissions. 
Objects cooler than about L0-L2 may exhibit outer-atmospheric
phenomena similar to those in the rotation-driven magnetosphere
of Jupiter rather than to those in convection-driven corona and
wind of the Sun.

The energetics of the main auroral oval of Jupiter dominate over the
phenomena associated with the lower-latitude structures that are
connected directly with the movement of Io, Europa, and Ganymede
through the jovian magnetic field, and those associated with the
higher-latitude polar-cap emissions that appear to be driven by the
solar wind. Jupiter's main auroral oval is not -~in contrast to Earth's~- the
separator between open and closed planetary magnetic field in the
interaction with the interplanetary magnetic field, but instead is
mapped to lower-latitude closed magnetic field (see, e.g., the
summarizing discussion by \citep{cowley+etal2003}). This auroral
structure is thought to be generated by the precipitation of energetic
electrons created by the electric current system that is involved in a
phenomenon referred to as "corotation" (or rather the breakdown
thereof) in the middle magnetosphere: beyond the interface where the
jovian magnetic field is strong enough to enforce corotation of the
plasma, a current-system is induced that includes a disk-shaped
near-equatorial extrusion in which field and plasma interact through
Lorentz forces to extract energy from the planet's rotation.

If the rotation of the coolest dwarf stars similarly powers their
magnetosphere, this energy is
ultimately drawn from the energy of rotation.  In the case of
the Jupiter, Io and Ganymede provide plasma 
conveniently high in the magnetosphere, outside the distance of
geosynchronous rotation but within the Alfv{\'e}n radius. In the case
of the ultra-cool dwarf stars, it is of course possible that one or
more close-in planets act as a similar plasma source, but rather than
postulating such planets, I hypothesize that plasma
is provided by a tenuous hot
stellar corona formed either by residual solar-like activity
(perhaps associated with the overturning field of a
turbulent dynamo) or magnetospheric activity 
(somehow formed by the breakdown of
corotation).  How, quantitatively, this
balance changes between L0 and L5 remains to be established.

For stars up to late-M, the power extracted from the stellar rotation
lies below the characteristic $H\alpha$ and X-ray emissions.  The
estimated rate $W_{\rm rot}$ at which rotational energy is lost from
an L2-type dwarf, in contrast, is larger than the characteristic
outer-atmospheric losses based on the coronal X-ray luminosity. In the
scenario of the rotationally-dominated magnetosphere,
\citet{eviatar+siscoe1980} argued that at most half of the total power
taken from Jupiter's rotational energy would be available for
radiative emissions, including the aurorae. The rough estimate of the
radiative losses of L-type dwarfs (as reflected by $L_X$) is not
inconsistent with that argument, but further observations are needed
to test this in more detail.

\referee{The quiescent radio emission from F--M type stars 
appears to be predominantly gyrosynchrotron
emission from high-energy coronal electrons
(see, e.g., the review by \citep{guedel2002}). 
\citet{hallinan+etal2008} argued that the radio signal
of several very cool dwarf stars (M8.5, M9,
and L3.5) is associated with the electron
cyclotron maser (ECM) process; they base that argument
on the observation that the radio signal is periodically
fully circularly polarized, whereas there is also 
an unpolarized component which they attribute to depolarized ECM
emission because its brightness temperature
is incompatible with incoherent synchrotron radiation.
They pointed out that the ECM mechanism is 
thought to be responsible for kHz--MHz emission
from Jupiter (cf., the review by \citep{zarka1998}).
Thus, both the enhanced radio to X-ray ratio 
and the radio polarization signature support
the analogy of the magnetospheric processes around Jupiter and around
ultracool radidly-rotating dwarf stars.}

The plasma transport in \referee{the magnetosphere
of Jupiter}, is thought to occur by convective
cells triggered by the centrifugal interchange instability, in which
cool, dense volumes change places with hotter, less dense volumes
(e.g., \citep{hill+etal2005}, \citep{chen+hill2008}). 
In the case of ultracool dwarf stars, the stellar equivalent of this
process into the stellar magnetotail shaped by the ISM wind could be
the process by which angular momentum is removed from
the star. 

In view of the arguments discussed in this paper, it appears warranted
to consider the possibility of a very similar system of currents in
the case of main-sequence stars cooler than about L0 as in the jovian
magnetosphere.  Perhaps much, if not most, of the H$\alpha$ emission
and the relatively weak coronal X-ray in stars cooler than L0 is
associated with activity (including aurorae) driven by the breakdown
of corotation like they occur on a smaller scale for Jupiter, while
the stellar radio emission may be a signature of high-energy particles
accelerated in that process. A dimensional analysis, developed by Fan
et al.\ (1982), applied to the case of L-type stars supports this
hypothesis quantitatively.

\acknowledgements I thank G.L.\ Siscoe, J.L.\ Linsky, and 
B.\ de Pontieu for helpful discussions and for their 
comments on the manuscript.

\begin{table}[t]
\caption{Characteristic stellar properties}\label{tab:table}
\begin{tabular}{ll|r||r|r}
\hline
Property & & M7V & dL2 & dL5 \\
\hline
Stellar radius &$R_\ast\,(R_\odot)$ & 0.16$^h$ & 0.09$^c$ & 0.08$^c$ \\
Effective temperature & $T_{\ast,eff}$ (K) & 2700$^a$ & 2080$^b$ & 1700$^b$ \\
Mass & $M_\ast\,(M_\odot)$ & 0.10$^h$ & 0.08$^c$ & 0.07$^c$ \\
Surface gravity & $\log{g}$ & 3.9 & 5.4 & 5.4 \\
Bolometric luminosity & $\log(L_{bol})$ & 30.7 & 29.7 & 29.3 \\
&&&&\\
Moment of inertia & $log(I)$ $^d$ & 51.4 & 50.8 & 50.6 \\
Equatorial rotation velocity & $v_{\rm eq}$ (km/s) & 10$^b$ & 20$^b$ & 30$^b$ \\
Rotational energy & $\log(W_{\rm rot})$ & 43.0 & 43.5 & 43.7 \\
Rotation period & $P$ (d) & 0.8 & 0.23 & 0.14 \\
Change in rotation period & $\dot{P}/P$ (Gyr) & 3$^e$ & 7$^e$ & $>10^e$ \\
Synchronous rotation &$R_S/R_\ast$&10.2&7.3&5.7\\
&&&&\\
Power from ISM wind & $\log(P_w)$ & 18.4 & 17.8& 17.7 \\
Relative luminosity in H$\alpha$ & $\log(L_{H\alpha}/L_{bol})$ & -4.3$^b$ & -5.7$^b$ & $<-7^b$ \\
Power in H$\alpha$ & $\log(L_{H\alpha})$ & 26.4 & 24.0 & $<22.3$ \\
Power in X-rays & $\log(L_X)$ & $\la 27.7^a$ & $\la 25.5^a$ & $\la 25.1^a$ \\
Excess of radio to X-ray & $\delta(L_R/L_X)$ $^f$ & $\sim \times 10$ &  $\sim \times 3000$ & ? \\
\\
Loss of rotation energy & $\log(W_{\rm rot})$ & 26.0 & 26.1 & $<26.2$ \\
Est. max. loss of rotation energy & $\log(W_{\rm C})$ & $\la$27.4 & $\la$25.9 & $\la$25.8\\
Power est. from Eq.~(\ref{eq:fan}) & $\log(W^\prime)$ & $24.9-25.9$ & $24.8-25.8$ & $25.1-26.1$ \\
\hline
\end{tabular}

$^a$ From Mohanty et al. (2004);\\
$^b$ Reiners and Basri (2008); West et al.\ (004);\\
$^c$ Baraffe et al. (2003), at age of 5\,Gy;\\
$^d$ For an approximate gyration radius of $0.3R_\ast$, cf.\ Claret and Gimenez (1990);\\
$^e$ Reiners and Basri (2008), their Fig.~9.\\
$^f$ Deviation from radio-X-ray relationship for warmer stars; 
Berger et al. (2008), their Fig.~9;\\
$^g$ Audard et al.\ (2007); \\
$^h$ from Allen (1972).
\end{table}
\nocite{baraffe+etal2003}
\nocite{claret+gimenez1990}
\nocite{allen72}
\nocite{west+etal2004}
\vfill\eject



\begin{thebibliography}{}

\bibitem[\protect\citeauthoryear{Allen}{1972}]{allen72}
Allen, C.~W.: 1972,
\newblock {\em {Astrophysical quantities}\/},
\newblock Athlone Press, Univ. of London, London, U. K.

\bibitem[\protect\citeauthoryear{{Audard} {\em
  et~al.\/}}{2007}]{audard+etal2007}
{Audard}, M., {Osten}, R.~A., {Brown}, A., {Briggs}, K.~R., {G{\"u}del}, M.,
  {Hodges-Kluck}, E., {\&} {Gizis}, J.~E.: 2007,
\newblock A{\&}A 471, L63

\bibitem[\protect\citeauthoryear{{Baraffe} {\em
  et~al.\/}}{2003}]{baraffe+etal2003}
{Baraffe}, I., {Chabrier}, G., {Barman}, T.~S., {Allard}, F., {\&}
  {Hauschildt}, P.~H.: 2003,
\newblock A{\&}A 402, 701

\bibitem[\protect\citeauthoryear{Basri}{2009}]{basri2008}
Basri, G.: 2009,
\newblock in {\em Cool Stars, Stellar Systems and the Sun 15\/}

\bibitem[\protect\citeauthoryear{{Berger}}{2006}]{berger2006}
{Berger}, E.: 2006,
\newblock ApJ 648, 629

\bibitem[\protect\citeauthoryear{{Berger} {\em
  et~al.\/}}{2008a}]{berger+etal2008}
{Berger}, E., {Basri}, G., {Gizis}, J.~E., {Giampapa}, M.~S., {Rutledge},
  R.~E., {Liebert}, J., {Mart{\'{\i}}n}, E., {Fleming}, T.~A., {Johns-Krull},
  C.~M., {Phan-Bao}, N., {\&} {Sherry}, W.~H.: 2008a,
\newblock ApJ 676, 1307

\bibitem[\protect\citeauthoryear{{Berger} {\em
  et~al.\/}}{2008b}]{berger+etal2008a}
{Berger}, E., {Rutledge}, R.~E., {Phan-Bao}, N., {Basri}, G., {Giampapa},
  M.~S., {Gizis}, J.~E., {Liebert}, J., {Martin}, E., {\&} {Fleming}, T.~A.:
  2008b,
\newblock ArXiv e-prints

\bibitem[\protect\citeauthoryear{{Chabrier} and
  {Baraffe}}{2000}]{chabrier+baraffe2000}
{Chabrier}, G. {\&} {Baraffe}, I.: 2000,
\newblock ARA{\&}A 38, 337

\bibitem[\protect\citeauthoryear{{Chen} and {Hill}}{2008}]{chen+hill2008}
{Chen}, Y. {\&} {Hill}, T.~W.: 2008,
\newblock Journal of Geophysical Research (Space Physics) 113(A12), 7215

\bibitem[\protect\citeauthoryear{{Claret} and
  {Gimenez}}{1990}]{claret+gimenez1990}
{Claret}, A. {\&} {Gimenez}, A.: 1990,
\newblock Ap{\&}SS 169, 215

\bibitem[\protect\citeauthoryear{{Cowley} {\em
  et~al.\/}}{2003}]{cowley+etal2003}
{Cowley}, S.~W.~H., {Bunce}, E.~J., {\&} {Nichols}, J.~D.: 2003,
\newblock Journal of Geophysical Research (Space Physics) 108, 8002

\bibitem[\protect\citeauthoryear{{Donati} {\em
  et~al.\/}}{2008}]{donati+etal2008}
{Donati}, J.-F., {Morin}, J., {Petit}, P., {Delfosse}, X., {Forveille}, T.,
  {Auri{\`e}re}, M., {Cabanac}, R., {Dintrans}, B., {Fares}, R., {Gastine}, T.,
  {Jardine}, M.~M., {Ligni{\`e}res}, F., {Paletou}, F., {Velez}, J.~C.~R., {\&}
  {Th{\'e}ado}, S.: 2008,
\newblock MNRAS 390, 545

\bibitem[\protect\citeauthoryear{{Eviatar} and
  {Siscoe}}{1980}]{eviatar+siscoe1980}
{Eviatar}, A. {\&} {Siscoe}, G.~L.: 1980,
\newblock Geophys. Res. Lett. 7, 1085

\bibitem[\protect\citeauthoryear{{Fan} {\em et~al.\/}}{1982}]{fan+etal1982}
{Fan}, C.~Y., {Hang}, H., {\&} {Wu}, J.: 1982,
\newblock ApJ 260, 353

\bibitem[\protect\citeauthoryear{{Feldman} {\em
  et~al.\/}}{1977}]{feldman+etal1977}
{Feldman}, W.~C., {Asbridge}, J.~R., {Bame}, S.~J., {\&} {Gosling}, J.~T.:
  1977,
\newblock in O.~R. {White} (Ed.), {\em The Solar Output and its Variation\/},
  p.~351

\bibitem[\protect\citeauthoryear{{Fleming} {\em
  et~al.\/}}{2003}]{fleming+etal2003}
{Fleming}, T.~A., {Giampapa}, M.~S., {\&} {Garza}, D.: 2003,
\newblock ApJ 594, 982

\bibitem[\protect\citeauthoryear{{G{\"u}del}}{2002}]{guedel2002}
{G{\"u}del}, M.: 2002,
\newblock ARA{\&}A 40, 217

\bibitem[\protect\citeauthoryear{{Hallinan} {\em
  et~al.\/}}{2006}]{hallinan+etal2006}
{Hallinan}, G., {Antonova}, A., {Doyle}, J.~G., {Bourke}, S., {Brisken}, W.~F.,
  {\&} {Golden}, A.: 2006,
\newblock ApJ 653, 690

\bibitem[\protect\citeauthoryear{{Hallinan} {\em
  et~al.\/}}{2008}]{hallinan+etal2008}
{Hallinan}, G., {Antonova}, A., {Doyle}, J.~G., {Bourke}, S., {Lane}, C., {\&}
  {Golden}, A.: 2008,
\newblock ApJ 684, 644

\bibitem[\protect\citeauthoryear{{Hallinan} {\em
  et~al.\/}}{2007}]{hallinan+etal2007}
{Hallinan}, G., {Bourke}, S., {Lane}, C., {Antonova}, A., {Zavala}, R.~T.,
  {Brisken}, W.~F., {Boyle}, R.~P., {Vrba}, F.~J., {Doyle}, J.~G., {\&}
  {Golden}, A.: 2007,
\newblock ApJL 663, L25

\bibitem[\protect\citeauthoryear{{Hansteen}}{2009}]{hpI-9}
{Hansteen}, V.~H.: 2009,
\newblock in C.~J. Schrijver and G.~L. Siscoe (Eds.), {\em Heliophysics I:
  Plasma physics of the local cosmos\/}, Cambridge University Press

\bibitem[\protect\citeauthoryear{{Hill} {\em et~al.\/}}{2005}]{hill+etal2005}
{Hill}, T.~W., {Rymer}, A.~M., {Burch}, J.~L., {Crary}, F.~J., {Young}, D.~T.,
  {Thomsen}, M.~F., {Delapp}, D., {Andr{\'e}}, N., {Coates}, A.~J., {\&}
  {Lewis}, G.~R.: 2005,
\newblock \grl 32, 14

\bibitem[\protect\citeauthoryear{{Liebert} {\em
  et~al.\/}}{1999}]{liebert+etal1999}
{Liebert}, J., {Kirkpatrick}, J.~D., {Reid}, I.~N., {\&} {Fisher}, M.~D.: 1999,
\newblock ApJ 519, 345

\bibitem[\protect\citeauthoryear{{Mohanty} and
  {Basri}}{2003}]{mohanti+basri2003}
{Mohanty}, S. {\&} {Basri}, G.: 2003,
\newblock ApJ 583, 451

\bibitem[\protect\citeauthoryear{{Mohanty} {\em
  et~al.\/}}{2002}]{mohanti+etal2002}
{Mohanty}, S., {Basri}, G., {Shu}, F., {Allard}, F., {\&} {Chabrier}, G.: 2002,
\newblock ApJ 571, 469

\bibitem[\protect\citeauthoryear{Reid {\em et~al.\/}}{1999}]{reid+etal99}
Reid, I.~N., Kirkpatrick, J.~D., Gizis, J.~E., {\&} Liebert, J.: 1999,
\newblock ApJL 527, 105

\bibitem[\protect\citeauthoryear{{Reiners}}{2007}]{reiners2007}
{Reiners}, A.: 2007,
\newblock Astronomische Nachrichten 328, 1040

\bibitem[\protect\citeauthoryear{{Reiners} and
  {Basri}}{2007}]{reiners+basri2007}
{Reiners}, A. {\&} {Basri}, G.: 2007,
\newblock ApJ 656, 1121

\bibitem[\protect\citeauthoryear{{Reiners} and
  {Basri}}{2008}]{reiners+basri2008}
{Reiners}, A. {\&} {Basri}, G.: 2008,
\newblock ApJ 684, 1390

\bibitem[\protect\citeauthoryear{{Reiners} and
  {Basri}}{2009}]{reiners+basri2009}
{Reiners}, A. {\&} {Basri}, G.: 2009,
\newblock ArXiv e-prints

\bibitem[\protect\citeauthoryear{{Robrade} and
  {Schmitt}}{2008}]{robrade+schmitt2008}
{Robrade}, J. {\&} {Schmitt}, J.~H.~M.~M.: 2008,
\newblock A{\&}A 487, 1139

\bibitem[\protect\citeauthoryear{Russell}{2007}]{russell2007}
Russell, C.~T.: 2007,
\newblock in V. Bothmer and I.~A. Daglis (Eds.), {\em Space weather: Physics
  and effects\/}, Springer, p.~103

\bibitem[\protect\citeauthoryear{{Schmidt} {\em
  et~al.\/}}{2007}]{schmidt+etal2007}
{Schmidt}, S.~J., {Cruz}, K.~L., {Bongiorno}, B.~J., {Liebert}, J., {\&}
  {Reid}, I.~N.: 2007,
\newblock AJ 133, 2258

\bibitem[\protect\citeauthoryear{Schrijver and Zwaan}{2000}]{schrijver+zwaan99}
Schrijver, C.~J. {\&} Zwaan, C.: 2000,
\newblock {\em {Solar and Stellar Magnetic Activity}\/},
\newblock Cambridge University Press, Cambridge, U.K.

\bibitem[\protect\citeauthoryear{{Walker} and
  {Russell}}{1995}]{walker+russell1995}
{Walker}, R.J. {\&} {Russell}, C.T.: 1995,
\newblock in M.G. {Kivelson} and C.T. {Russell} (Eds.), {\em Introduction to
  Space Physics\/}, Cambridge University Press

\bibitem[\protect\citeauthoryear{{West} {\em et~al.\/}}{2004}]{west+etal2004}
{West}, A.~A., {Hawley}, S.~L., {Walkowicz}, L.~M., {Covey}, K.~R.,
  {Silvestri}, N.~M., {Raymond}, S.~N., {Harris}, H.~C., {Munn}, J.~A.,
  {McGehee}, P.~M., {Ivezi{\'c}}, {\v Z}., {\&} {Brinkmann}, J.: 2004,
\newblock AJ 128, 426

\bibitem[\protect\citeauthoryear{{Wood} {\em et~al.\/}}{2002}]{wood+etal2002}
{Wood}, B.~E., {M{\"u}ller}, H., {Zank}, G.~P., {\&} {Linsky}, J.~L.: 2002,
\newblock ApJ 574, 412

\bibitem[\protect\citeauthoryear{{Wood} {\em et~al.\/}}{2005}]{wood+etal2005}
{Wood}, B.~E., {M{\"u}ller}, H.-R., {Zank}, G.~P., {Linsky}, J.~L., {\&}
  {Redfield}, S.: 2005,
\newblock ApJL 628, L143

\bibitem[\protect\citeauthoryear{{Zarka}}{1998}]{zarka1998}
{Zarka}, P.: 1998,
\newblock JGR 103, 20159

\end{thebibliography}
\end{document}